\documentclass[aps,prd,preprint,floats,superscriptaddress]{revtex4}
\usepackage{graphicx}

\def\ie{{\it i.e.}}

\def\re{{\rm Re}}

\def\be{\begin{equation}}
\def\ee{\end{equation}}
\def\bea{\begin{eqnarray}}
\def\eea{\end{eqnarray}}
\def\bean{\begin{eqnarray*}}
\def\eean{\end{eqnarray*}}
\def\bary{\begin{array}}
\def\eary{\end{array}}
\def\nn{\nonumber}

\def\lan{\langle}
\def\ran{\rangle}

\def\lbar{\overline}

\begin{document}

\preprint{
  \vbox{
    \hbox{MADPH-04-1380; ANL-HEP-PR-04-39;}
    \hbox{UPR-1078T; hep-ph/0405108}
  }}

\title{$B_s$-$\lbar B_s$ Mixing in $Z'$ Models with Flavor-Changing Neutral
  Currents}

\author{Vernon Barger}
\email[e-mail: ]{barger@oriole.physics.wisc.edu}
\affiliation{Department of Physics, University of Wisconsin, Madison, WI 53706}
\author{Cheng-Wei Chiang}
\email[e-mail: ]{chengwei@physics.wisc.edu}
\affiliation{Department of Physics, University of Wisconsin, Madison, WI 53706}
\author{Jing Jiang}
\email[e-mail: ]{jiangj@hep.anl.gov}
\affiliation{HEP Division, Argonne National Laboratory,
9700 S. Cass Avenue, Argonne, IL 60439}
\author{Paul Langacker}
\email[e-mail: ]{pgl@electroweak.hep.upenn.edu}
\affiliation{Department of Physics and Astronomy,
University of Pennsylvania, Philadelphia, PA 19104-6396}

\date{\today}

\begin{abstract}
  In models with an extra $U(1)'$ gauge boson family non-universal couplings to
  the weak eigenstates of the standard model fermions generally induce
  flavor-changing neutral currents.  This phenomenon leads to interesting
  results in various $B$ meson decays, for which recent data indicate hints of
  new physics involving significant contributions from $b \to s$ transitions.
  We analyze the $B_s$ system, emphasizing the effects of a $Z'$ on the mass
  difference and $CP$ asymmetries.
\end{abstract}

%11.30.Er  Charge conjugation, parity, time reversal, and other discrete
%          symmetries 
%11.30.Hv  Flavor symmetries
%12.15.Hh  Determination of Kobayashi-Maskawa matrix elements
%13.25.Hw  Decays of bottom mesons
%14.40.Nd  Bottom mesons

\maketitle

%%%%%%%%%%%%%%%%%%%%%%%%%%%%%%%%%%%%%%%%%%%%%%%%%%%%%%%%%%%%

%%%%%%%%%%%%%%%%%%%%%%%%%%%%%%%%%%%%%%%%
\section{Introduction \label{sec:intro}}
%%%%%%%%%%%%%%%%%%%%%%%%%%%%%%%%%%%%%%%%

The study of $B$ physics and the associated CP violating observables has been
suggested as a good means to extract information on new physics at low energy
scales \cite{Grossman:1996ke,Fleischer:1996bv,London:1997zk,Fleischer:2001pc,%
  Hiller:2002ci,Ciuchini:2002pd,Neubert:SuperB}.  Since $B$-$\lbar B$ mixing is
a loop-mediated process within the standard model (SM), it offers an
opportunity to see the footprints of physics beyond the SM.  The currently
observed $\Delta M_d = 0.489 \pm 0.008$ ps$^{-1}$ \cite{Hagiwara:fs} and its
mixing phase $\sin 2\beta = 0.736 \pm 0.049$ extracted from the $B_d \to J/\psi
K_S$ mode \cite{Browder:2003ii} agree well with constraints obtained from other
experiments \cite{Hocker:2001xe}.  However, no such information other than a
lower bound $\Delta M_s > 14.5$ ps$^{-1}$ \cite{HFAG} is available for the
$B_s$ meson yet.

Based upon SM predictions, $\Delta M_{B_s}$ is expected to be about $18$
ps$^{-1}$ and its mixing phase $\phi_s$ only a couple of degrees.  In contrast
to the $B_d$ system, the more than 25 times larger oscillation frequency and a
factor of four lower hadronization rate from $b$ quarks pose the primary
challenges in the study of $B_s$ oscillation and $CP$ asymmetries.  Since the
$B_s \to J/\psi \phi$ decay is dominated by a Cabibbo-Kobayashi-Maskawa (CKM)
favored tree-level process, $b \to c {\bar c} s$, that does not involve a
different weak phase in the SM, its asymmetry provides the most reliable
information about the mixing phase $\phi_s$.

Although new physics contributions may not compete with the SM processes in
most of the $b \to c {\bar c} s$ decays, they can play an important role in
$B_s$-$\lbar B_s$ mixing because of its loop nature in the SM \cite{NPrefs}.
In particular, the mixing can be significantly modified in models in which a
tree-level $b\to s$ transition is present.  Thus, measurement of the properties
of $B_s$ meson mixing is of high interest in future $B$ physics studies as a
means to reveal new physics \cite{Grossman:1997pa,Dunietz:2000cr}.  Since the
current $B$ factories do not run at the $\Upsilon(5S)$ resonance to produce
$B_s$ mesons, it is one of the primary objectives of hadronic colliders to
study $B_s$ oscillation and decay in the coming years
\cite{Anikeev:2001rk,Ball:2000ba}.

Flavor changing neutral currents (FCNC) coupled to an extra $U(1)'$ gauge boson
arise when the $Z'$ couplings to physical fermion eigenstates are non-diagonal.
One way for this to happen is by the introduction of exotic fermions with
different $U(1)'$ charges that mix with the SM fermions
\cite{Nardi:1992nq,Babu:1996vt,Rizzo:1998ut,Leroux:2001fx,Langacker:2000ju} as
occurs in $E_6$ models.  In the $E_6$ case, mixing of the right-handed ordinary
and exotic quarks, all $SU(2)_L$ singlets, induces FCNC mediated by a heavy
$Z'$ or by (small) $Z$-$Z'$ mixing, so the quark mixing can be large.  Mixing
between ordinary (doublet) and exotic (singlet) left-handed quarks induces FCNC
mediated by the SM $Z$ boson \cite{Leroux:2001fx}.  We will also allow for this
possibility, but in this case the quark mixing must be very small.

Another possibility involves family non-universal couplings.  It is well-known
that string models naturally give extra $U(1)'$ groups, at least one of which
has family non-universal couplings to the SM fermions
\cite{Chaudhuri:1994cd,Cleaver:1998gc,Cvetic:2001tj,Cvetic:2002qa}.
Generically, the physical and gauge eigenstates do not coincide.  Here, unlike
the above-mentioned $E_6$ case, off-diagonal couplings of fermions to the $Z'$
boson can be obtained without mixing with additional fermion states.  In these
types of models, both left-handed and right-handed fermions can have family
non-diagonal couplings with the $Z'$, while couplings to the $Z$ are family
diagonal (up to small effects from $Z-Z'$ mixing).

The $Z'$ contributions to $B_s$-$\lbar B_s$ mixing are related to those for
hadronic, semileptonic, and leptonic $B$ decays in specific models in which the
diagonal $Z'$ couplings to $q \bar q$, $\ell^+ \ell^-$, etc. are known, but are
independent in general \footnote{$B_s$-$\lbar B_s$ mixing in leptophobic $E_6$
  models was considered in Ref.~\cite{Leroux:2001fx}.  A model with
  non-universal right-handed couplings was discussed in~\cite{He:2004it}.
  Mechanisms for flavor change in dynamical symmetry breaking models are
  described in~\cite{dsb}.}.  We have found that in specific models,
$B_s$-$\lbar B_s$ mixing effects can be significant while being consistent with
the other constraints; these results will be presented elsewhere.  In the
present paper, we will treat the mixing in a model-independent way.

Recently, we have studied the implications of a sizeable off-diagonal $Z'$
coupling between the bottom and strange quark in the indirect CP asymmetry of
$B \to \phi K_S$ decay \cite{Barger:2003hg}, which appears to show a
significant deviation from the SM prediction
\cite{Grossman:1997gr,Hiller:2002ci,Ciuchini:2002pd,Chiang:2003jn}.  Here we
extend our analysis to $B_s$-$\lbar B_s$ mixing where the $Z'$ contributions
also enter at the tree level.  Applications to the $B \to \pi K$ anomaly are
under investigation \cite{BCLL}.

The paper is organized as follows.  In Section \ref{sec:mixing}, we review the
basic formalism of $B_s$-$\lbar B_s$ mixing.  In Section \ref{sec:sm}, we
evaluate $\Delta M_s$ in the SM.  In Section \ref{sec:zpr}, we include the $Z'$
contributions, allowing both left-handed and right-handed couplings in the
mixing, and study their effects on observables.  Our main results are
summarized in Section \ref{sec:conclusion}.

%%%%%%%%%%%%%%%%%%%%%%%%%%%%%%%%%%%%%%%%
\section{$B_s$-$\lbar B_s$ Mixing \label{sec:mixing}}
%%%%%%%%%%%%%%%%%%%%%%%%%%%%%%%%%%%%%%%%

In the conventional decomposition of the heavy and light eigenstates
\bea
|B_s \rangle_L = p |B_s^0 \rangle + q |\lbar B_s^0 \rangle ~, \nn \\
|B_s \rangle_H = p |B_s^0 \rangle - q |\lbar B_s^0 \rangle ~,
\eea
the mixing factor
\be
\label{eq:mix}
\left(\frac{q}{p}\right)_{\rm SM}
\simeq \sqrt{\frac{M_{12}^{\rm SM*}}{M_{12}^{\rm SM}}} ~,
\ee
has a phase
\be
\phi_s^{SM}
= 2 \arg(V_{tb}V_{ts}^*)
= -2\lambda^2{\bar \eta}
\simeq -2^{\circ} ~,
\qquad \sin2\phi_s^{SM} \simeq -0.07 ~,
\label{eq:15}
\ee
where the theoretical expectation $\Gamma_{12}^{\rm SM} \ll M_{12}^{\rm SM}$ is
used.  The approximate formula Eq.~(\ref{eq:mix}) receives a small correction
once $\Gamma_{12}^{\rm SM}$ is included.  Model independently, this only shifts
$\phi_s$ is at the few percent level.  With errors on $\lambda$ and ${\bar
  \eta}$ included, we have the SM expectation that $\sin2\phi_s^{SM} \simeq
-0.07 \pm 0.01$.

The off-diagonal element of the decay matrix, $\Gamma_{12}^{\rm SM}$, is
evaluated by considering decay channels that are common to both $B_s$ and
$\lbar B_s$ mesons, and $M_{12}$ is the off-diagonal element of the mass
matrix.  Due to the CKM enhancement, $\Gamma_{12}^{\rm SM}$ is dominated by the
charm-quark contributions over the up-quark contribution in a box diagram.
Unlike the Kaon system, $\Gamma_{12}^{\rm SM}$ is much smaller than
$M_{12}^{\rm SM}$ for $B$ mesons because the former is related to the $B$ meson
decays and set by the scale of its mass, whereas the latter is proportional to
$m_t^2$.  We can safely assume that $\Gamma_{12}$ is not significantly modified
by new physics because $\Gamma_{12}$ receives major contributions from CKM
favored $b \to c {\bar c} s$ decays in the SM, and the SM result $\Gamma_{12}
\ll M_{12}$ is unlikely to change.

The mass difference of the two physical states is
\be
\Delta M_s \equiv M_H - M_L
\simeq 2 |M_{12}| ~.
\ee
The width difference is
\be
\label{eq:widthdiff}
\Delta \Gamma
\equiv \Gamma_H - \Gamma_L
= \frac{2 \re(M_{12}^* \Gamma_{12})}{|M_{12}|}
= 2 |\Gamma_{12}| \cos\theta ~,
\ee
where the relative phase is $\theta = \arg (M_{12} / \Gamma_{12})$.  Since
$\Gamma_{12}$ is dominated by the contributions from CKM favored $b \to c {\bar
  c} s$ decays, we have $\theta = \arg \left( (V_{tb} V_{ts}^*) / (V_{cb}
  V_{cs}^*) \right) \simeq \pi$, and thus $\Delta \Gamma \simeq - 2
|\Gamma_{12}|$ 
is negative in the SM.  Although $\Gamma_{12}$ is unlikely to be affected by
new physics, the width difference always increases as long as the weak phase of
$M_{12}$ gets modified \cite{Grossman:1996er}.

The observability of $B_s$-$\lbar B_s$ oscillations is often indicated by the
parameter
\begin{equation}
  x_s
  \equiv \frac{\Delta M_s}{\Gamma_s} ~,
\end{equation}
where $\Gamma_s = (4.51 \pm 0.18) \times 10^{-13}$ GeV, converted from the
world average lifetime $\tau_s = 1.461 \pm 0.057$ ps \cite{Hagiwara:fs}.  The
expected large value of $x_s$ is a challenge for experimental searches.
Currently, the result from all ALEPH \cite{Heister:2002gk}, CDF
\cite{Abe:1998qj}, DELPHI \cite{Adam:1997pv}, OPAL \cite{Abbiendi:1999gm}, and
SLD \cite{Abe:2002ua} studies of $\Delta M_s$ with a combined $95\%$ confidence
level (CL) sensitivity on $\Delta M_s$ of $18.3$ ps$^{-1}$ gives \cite{HFAG}
\be
  \Delta M_s > 14.5 \, {\rm ps}^{-1} ~, \quad {\rm and} \quad
  x_s > 20.8 ~.
\ee
It is also measured that $m_{B_s} = 5369.6 \pm 2.4$ MeV \cite{Hagiwara:fs} and
$\Delta \Gamma_s/\Gamma_s = -0.16^{+0.15}_{-0.16} (|\Delta \Gamma_s|/\Gamma_s<
0.54)$ (with the $95\%$ CL upper bound given in parentheses \cite{HFAG})
consistent with recent next-to-leading-order (NLO) QCD estimates
\cite{Beneke:1998sy}.  In comparison, the $B_d$ system has $m_{B_d} = 5279.4
\pm 0.5$ MeV, $\Delta M_d = (0.489 \pm 0.008)$ ps$^{-1}$, $x_d = 0.755 \pm
0.015$, and $\tau_{B_d} = 1.542 \pm 0.076$ ps \cite{Hagiwara:fs}.

%%%%%%%%%%%%%%%%%%%%%%%%%%%%%%%%%%%%%%%%
\section{$\Delta M_s$ in the SM \label{sec:sm}}
%%%%%%%%%%%%%%%%%%%%%%%%%%%%%%%%%%%%%%%%

The $|\Delta B| = 2$ and $|\Delta S| = 2$ operators relevant for our
discussions are:
\begin{eqnarray}
  && O^{LL}
  = [\bar s \gamma_{\mu} (1 - \gamma_5) b]
    [\bar s \gamma^{\mu} (1 - \gamma_5) b] ~, \nn \\
  && O_1^{LR}
  = [\bar s \gamma_{\mu} (1 - \gamma_5) b]
    [\bar s \gamma^{\mu} (1 + \gamma_5) b] ~, \nn \\
  && O_2^{LR}
  = [\bar s (1 - \gamma_5) b] [\bar s (1 + \gamma_5) b] ~, \nn \\
  \label{eq:ops}
  && O^{RR}
  = [\bar s \gamma_{\mu} (1 + \gamma_5) b]
    [\bar s \gamma^{\mu} (1 + \gamma_5) b] ~.
\end{eqnarray}
Because of the $V-A$ structure, only the operator $O^{LL}$ contributes to
$B_s$-$\lbar B_s$ mixing in the SM.  The other three operators appear in the
$Z'$ models because of the right-handed couplings and operator mixing through
renormalization, as considered in the next section.

In the SM the contributions to
\be
M_{12}^{\rm SM} \simeq
  \frac{1}{2m_{B_s}}
  \langle B_s^0 | {\cal H}_{\rm eff}^{\rm SM} | \lbar B_s^0 \rangle
\ee
are dominated by the top quark loop.  The result, accurate to NLO in
QCD, is given by \cite{Buchalla:1995vs} 
\begin{equation}
  M_{12}^{\rm SM}
  = \frac{G_F^2}{12\pi^2} M_W^2 m_{B_s} f_{B_s}^2 (V_{tb}V_{ts}^*)^2
    \eta_{2B} S_0(x_t) [\alpha_s(m_b)]^{-6/23}
    \left[ 1 + \frac{\alpha_s(m_b)} {4\pi} J_5 \right] B^{LL}(m_b)
    ~,
\label{eq:m12}
\end{equation}
where $x_t=(m_t(m_t)/M_W)^2$ and
\begin{equation}
  S_0(x)
  = \frac{4x - 11x^2 + x^3}{4(1-x)^2} - \frac{3x^3\ln x}{2(1 - x)^3} ~.
\end{equation}
Using $m_t(m_t) = 170 \pm 5$ GeV, we find $S_0(x_t) = 2.463$.  The NLO
short-distance QCD corrections are encoded in the parameters $\eta_{2B} \simeq
0.551$ and $J_5 \simeq 1.627$ \cite{Buchalla:1995vs}.  The bag parameter
$B^{LL}(\mu)$ is defined through the relation
\begin{equation}
  \lan \lbar B_s | O^{LL} | B_s \ran
  \equiv \frac83 m_{B_s}^2 f_{B_s}^2 B^{LL}(\mu) ~.
\end{equation}
In the following numerical analysis,
we will use $G_F = 1.16639 \times 10^{-5} \,\,{\rm GeV}^{-2}$ and $M_W = 80.423
\pm 0.039$ GeV \cite{Hagiwara:fs}, and write the SM part of $\Delta M_s$ as
\begin{equation}
 \Delta M_s^{\rm SM}
 = 1.19 \left| \frac{V_{tb}V_{ts}^*}{0.04} \right|^2
   \left(\frac{f_{B_s}}{230 \; {\rm MeV}}\right)^2
   \left(\frac{B^{LL}(m_b)}{0.872}\right) \times 10^{-11}~{\rm GeV} ~.
\label{eq:deltam}
\end{equation}

Current lattice calculations still show quite large errors on the hadronic
parameters $f_{B_s} = 230 \pm 30$ MeV and $B^{LL}(m_b) = 0.872 \pm 0.005$
\cite{Yamada:2001xp,Becirevic:2001xt,Ryan:2001ej}.  However, the ratio
\begin{equation}
  \label{eq:xi}
  \xi \equiv
  \frac{f_{B_s} \sqrt{{\hat B}_{B_s}}}{f_{B_d} \sqrt{{\hat B}_{B_d}}}
\end{equation}
can be determined with a much smaller theoretical error, where ${\hat B}_{B_q}$
is the renomalization-independent bag parameter for the $B_q$ meson ($q=d,s$).
Therefore, the error on $\Delta M_s$ within the SM can be evaluated by
comparing with $\Delta M_d$, \ie,
\begin{equation}
  \Delta M_s^{\rm SM} =
  \Delta M_d^{\rm SM} \, \xi \,\frac{m_{B_s}}{m_{B_d}}
  \frac{(1-\lambda^2)^2}{\lambda^2 \left[(1-\lbar\rho)^2 +
  {\lbar\eta}^2\right]} ~.
\end{equation}
Using the measured values of the Wolfenstein parameters
\cite{Wolfenstein:1983yz} $\lambda = 0.2265 \pm 0.0024$, $A = 0.801 \pm 0.025$,
$\lbar\rho = 0.189 \pm 0.079$, and $\lbar\eta = 0.358 \pm 0.044$
\cite{Hocker:2001xe}, $\xi = 1.24 \pm 0.07$ \cite{Battaglia:2003in}, and the
mass parameters quoted above, we obtain the SM predictions
\begin{eqnarray}
  \Delta M_s^{\rm SM}
  &=& (1.19 \pm 0.24) \times 10^{-11}~{\rm GeV} \nn \\
  &=& 18.0 \pm 3.7~{\rm ps}^{-1} ~, \nn \\
  x_s^{\rm SM} &=& 26.3 \pm 5.5 ~.
\end{eqnarray}
As noted above, the central value of $x_s$ is slightly larger than the current
sensitivity based upon the world average.  Recent LHC studies show that with
one year of data, $\Delta M_s$ can be explored up to 30 ps$^{-1}$ (ATLAS), 26
ps$^{-1}$ (CMS), and 48 ps$^{-1}$ (LHCb) (corresponding to $x_s$ up to 46, 42,
and 75); the LHCb result is based on exclusive hadronic decay modes
\cite{Ball:2000ba}.  The sensitivity of both CDF and BTeV on $x_s$ can also
reach up to 75 using the same modes \cite{Anikeev:2001rk}, for a luminosity of
2 fb$^{-1}$.  The sensitivity on $\sin2\phi_s$ is correlated with the value of
$x_s$, and it becomes worse as $x_s$ increases.  A statistical error of a few
times $10^{-2}$ can be reached at CMS and LHCb for moderate $x_s \simeq 40$
\cite{Ball:2000ba}.

%%%%%%%%%%%%%%%%%%%%%%%%%%%%%%%%%%%%%%%%
\section{$Z'$ Contributions \label{sec:zpr}}
%%%%%%%%%%%%%%%%%%%%%%%%%%%%%%%%%%%%%%%%

For simplicity, we assume that there is no mixing between the SM $Z$ and the
$Z'$ (small mixing effects can be easily incorporated \cite{Langacker:2000ju}).
A purely left-handed off-diagonal $Z'$ coupling to $b$ and $s$ quarks results
in an effective $|\Delta B| = 2$, $|\Delta S| = 2$ Hamiltonian at the $M_W$
scale of
\begin{equation}
  \label{eq:para}
  {\cal H_{\rm eff}^{Z'}} = \frac{G_F}{\sqrt{2}} \left( \frac{g_2
      M_Z}{g_1 M_{Z'}} B_{sb}^L \right)^2  O^{LL}(m_b) \equiv 
  \frac{G_F}{\sqrt{2}} \rho_L^2 e^{2 i \phi_L} O^{LL}(m_b)~,
\end{equation}
where $g_2$ is the $U(1)'$ gauge coupling, $g_1 = e / (\sin\theta_W
\cos\theta_W)$, $M_{Z'}$ is the mass of the $Z'$, and $B_{sb}^L$ is the FCNC
$Z'$ coupling to the bottom and strange quarks.  The parameters $\rho_L$ and
the weak phase $\phi_L$ in the $Z'$ model are defined by the second equality.
Generically, we expect that $g_2/g_1 \sim 1$ if both $U(1)$ groups have the
same origin from some grand unified theory, and $M_Z/M_{Z'} \sim 0.1$ for a
TeV-scale $Z'$.  If $|B_{sb}^L| \sim |V_{tb} V_{ts}^*|$, then an
order-of-magnitude estimate gives us $\rho_L \sim {\cal O}(10^{-3})$, which is
in the ballpark of giving significant contributions to the $B_s$-$\lbar B_s$
mixing.  The $Z'$ does not contribute to $\Gamma_{12}$ at tree level because
the intermediate $Z'$ cannot be on shell.  After evolving from the $M_W$ scale
to $m_b$, the effective Hamiltonian becomes \cite{Buchalla:1995vs}
\begin{equation}
  {\cal H_{\rm eff}^{Z'}}
  = \frac{G_F}{\sqrt{2}} \left[ 1 +
    \frac{\alpha_s(m_b)-\alpha_s(M_W)}{4\pi} J_5\right]
  R^{6/23} \rho_L^2 e^{2 i \phi_L} O^{LL}(m_b) ~,
\label{eq:eHsm}
\end{equation}
where $R = \alpha_s(M_W) / \alpha_s(m_b)$.  Although the above effective
Hamiltonian is largely suppressed by the ratio $(g_2 M_Z)/(g_1 M_{Z'})$, it
contains only one power of $G_F$ in comparison with the corresponding quadratic
dependence in the SM because the $Z'$-mediated process occurs at tree level.

The full description of the running of the Wilson coefficient from the $M_W$
scale to $m_b$ can be found in \cite{Buchalla:1995vs}.  We only repeat the
directly relevant steps here.  The renormalization group equation for the
Wilson coefficients $\vec{C}$,
\begin{equation}
  \frac{d}{d\,\ln\mu} \vec{C} = \gamma^T(g) \vec{C}(\mu) ~,
\end{equation}
can be solved with the help of the $U$ matrix
\begin{equation}
  \vec{C}(\mu) = U(\mu,M_W) \vec{C}(M_W) ~,
\end{equation}
in which $\gamma^T(g)$ is the transpose of the anomalous dimension matrix
$\gamma(g)$.  With the help of $d g/d \ln\mu = \beta(g)$, $U$ obeys the same
equation as $\vec{C}(\mu)$.  We expand $\gamma(g)$ to the first two terms in
the perturbative expansion,
\begin{equation}
  \gamma(\alpha_s)
  = \gamma^{(0)} \frac{\alpha_s}{4\pi} +
  \gamma^{(1)}\left(\frac{\alpha_s}{4\pi}\right)^2 ~.
\end{equation}
To this order the evolution matrix $U(\mu, m)$ is given by
\begin{equation}
  U(\mu,m)=
  \left(1+{\alpha_s(\mu)\over 4\pi} J\right) U^{(0)}(\mu,m)
  \left(1-{\alpha_s(m)\over 4\pi} J\right) 
  ~,
\end{equation}
where $U^{(0)}$ is the evolution matrix in leading logarithmic approximation
and the matrix $J$ expresses the next-to-leading corrections.  We have
\begin{equation}
\label{u0vd}
  U^{(0)}(\mu,m)
  = V \left(
  {\left[{\alpha_s(m)\over\alpha_s(\mu)}\right]}^{{\vec\gamma^{(0)}\over
  2\beta_0}} \right)_D V^{-1} ~,
\end{equation}
where $V$ diagonalizes ${\gamma^{(0)T}}$, \ie, $\gamma^{(0)}_D=V^{-1}
{\gamma^{(0)T}} V$, and $\vec\gamma^{(0)}$ is the vector containing the
diagonal elements of the diagonal matrix $\gamma^{(0)}_D$.  In terms of
$G=V^{-1} {\gamma^{(1)T}} V$ and a matrix $H$ whose elements are
\begin{equation}
\label{jvs}
  H_{ij}
  = \delta_{ij}\gamma^{(0)}_i{\beta_1\over 2\beta^2_0} -
    {G_{ij}\over 2\beta_0+\gamma^{(0)}_i-\gamma^{(0)}_j} ~,
\end{equation}
the matrix $J$ is given by $J=V H V^{-1}$.

The operators $O^{LL}$ and $O^{RR}$ do not mix with others under
renormalization.  Their Wilson coefficients follow exactly the same RGEs, where
the above-mentioned matrices are all simple numbers.  The factor
\begin{equation}
  \left[ 1 + \frac{\alpha_s(m_b)-\alpha_s(M_W)}{4\pi} J_5\right] R^{6/23}
\end{equation}
in Eq.~(\ref{eq:eHsm}) reflects the RGE running.  On the other hand, $O_1^{LR}$
and $O_2^{LR}$ form a sector that is mixed under RG running.  Although the $Z'$
boson only induces the operator $O_1^{LR}$ at high energy scales, $O_2^{LR}$ is
generated after evolution down to low energy scales and, in particular, its
Wilson coefficient $C_2^{LR}$ is strongly enhanced by the RG effects
\cite{Buras:2000if}.

With contributions from both the SM and the $Z'$ boson with only left-handed
FCNC couplings included, the $B_s$ mass difference is
\begin{eqnarray}
 \Delta M_s &=& \Delta M_s^{\rm SM} \left( 1 + \frac{\Delta M_s^{Z'}}{\Delta
     M_s^{\rm SM}} \right)
 = 18.0 \left| 1 + 3.858 \times 10^5 \rho_L^2 e^{2 i \phi_L}
 \right| \; {\rm ps}^{-1}~,
\end{eqnarray}
The corresponding result for the oscillation parameter is
\begin{equation}
  \label{eq:xsL}
  x_s
  = 26.3 \left| 1 + 3.858 \times10^5 \rho_L^2 e^{2 i \phi_L} \right| ~.
\end{equation}
With couplings of only one chirality, the physical observables $\Delta M_s$,
$x_s$, and $\sin 2\phi_s$ are periodic functions of the new weak phase $\phi_L$
with a period of $180^{\circ}$.

%Figure 1
\begin{figure}[t]
\centerline{\includegraphics[width=8cm]{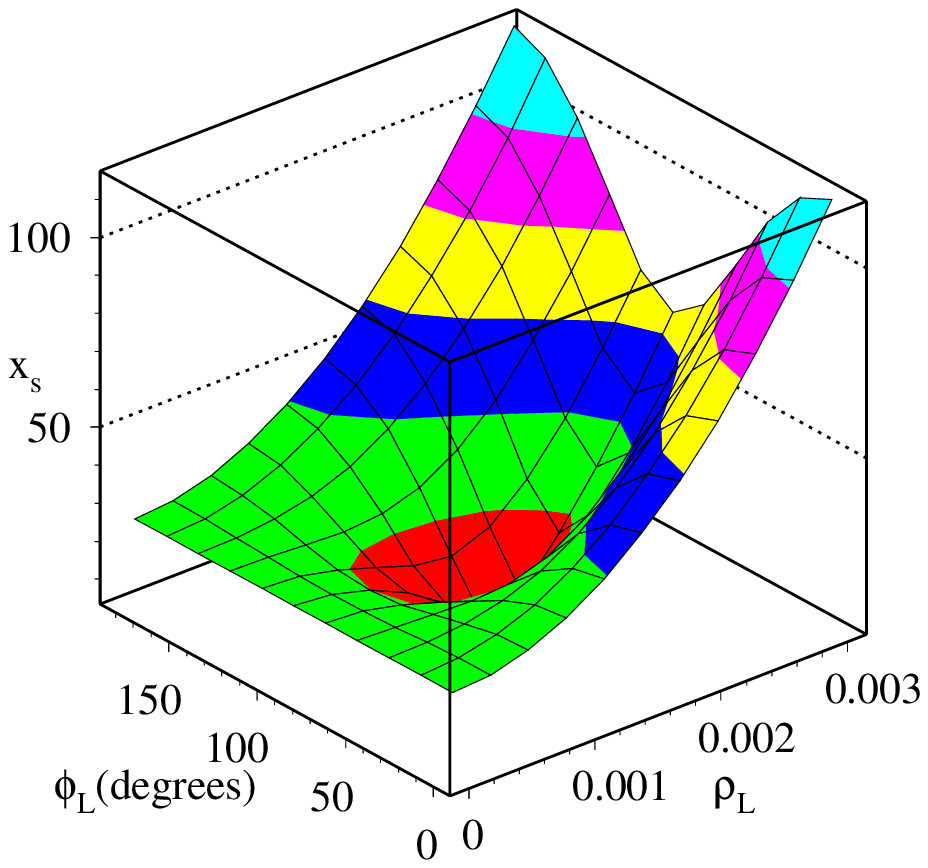}
            \includegraphics[width=8cm]{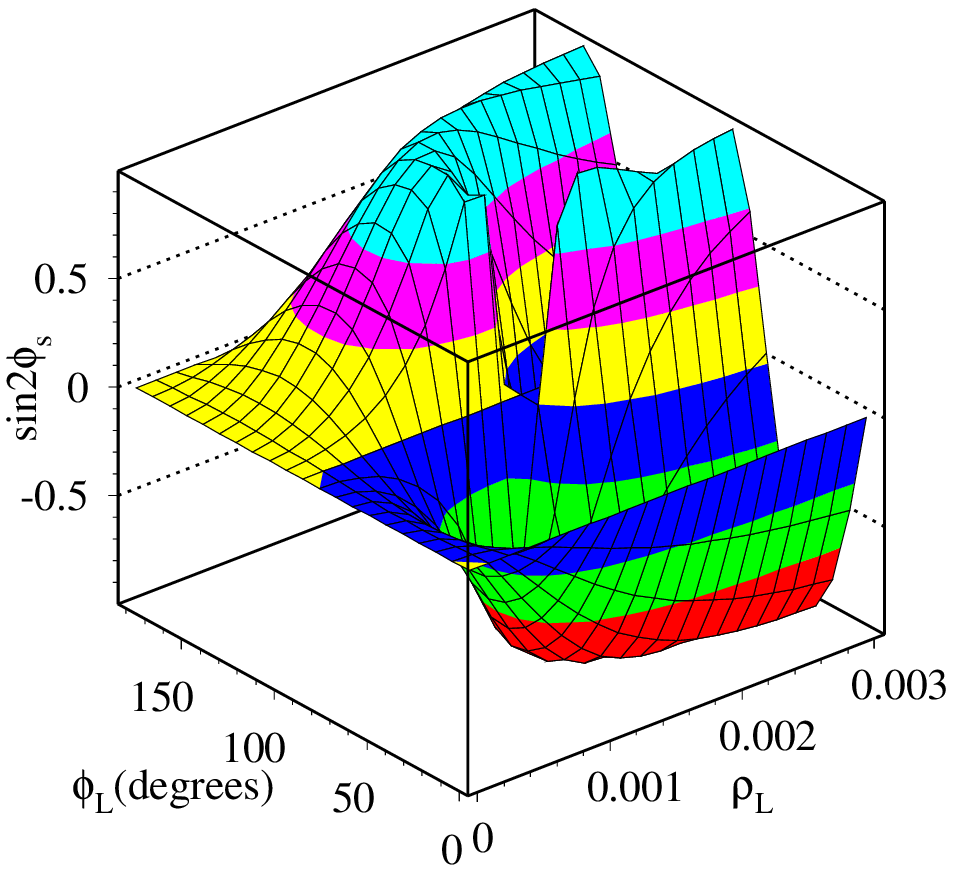}}
\centerline{(a)\hspace{8cm}(b)}
\caption[]{Three-dimensional plot of $x_s$ (a) and $\sin 2\phi_s$ (b)
  versus $\rho_L$ and $\phi_L$ with a $Z'$-mediated FCNC for
  left-handed $b$ and $s$ quarks.  The color shadings in both plots
  have no specific physical meaning. 
\label{fig:ll3d}}
\end{figure}

Fig.~\ref{fig:ll3d} (a) shows the effects of including a $Z'$ with left-handed
coupling.  We see that if $\rho_L$ is small, $x_s$ is dominated by the SM
contribution and has a value $\sim 26$.  For $\phi_L$ around $90^\circ$ and
$\rho_L$ between $0.001$ and $0.002$, the $Z'$ contribution tends to cancel
that of the SM and reduces $x_s$ to be smaller than the SM value of $26.3$.  In
Eq.~(\ref{eq:xsL}) and Fig.~\ref{fig:ll3d} (a), we see that the $Z'$ has a
comparable contribution to the SM if $\rho_L \agt 0.002$, independent of the
actual value of $\phi_L$. The planned resolution of Fermilab Run II and LHCb
are both about $x_s \alt 75$ \cite{Anikeev:2001rk,Ball:2000ba}.  Thus, a
$\rho_L$ greater than about $0.003$ will result in an $x_s$ beyond the planned
sensitivity. If $x_s$ is measured to fall within a range, one can read from the
plot what the allowed region is for the chiral $Z'$-model parameters. The same
discussion can easily be applied to a $Z'$ model with only right-handed
couplings.  Fig.~\ref{fig:ll3d} (b) shows $\sin 2\phi_s$ as a function of
$\rho_L$ and $\phi_L$. As $\rho_L$ increases, $\sin 2\phi_s$ goes through more
oscillations when $\phi_L$ varies from $0$ to $\pi$.

%FIgure 2
\begin{figure}[t]
\centerline{\includegraphics[width=13cm]{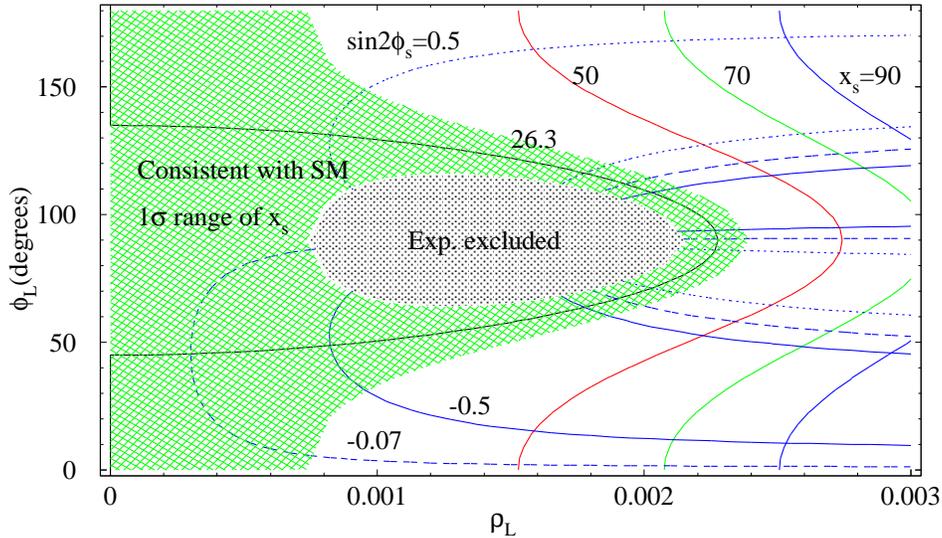}}
\caption[]{Contour plot of $x_s$ and $\sin 2\phi_s$ with a
  $Z'$-mediated FCNC for left-handed $b$ and $s$ quarks.  The shaded region is
  for $x_s < 20.6$, which is excluded at $95\%$ CL by experiments.  The hatched
  region corresponds to $1 \sigma$ ranges around the SM value $x_s = 26.3 \pm
  5.5$ (black curve).  The solid curves open to the left are contours for $x_s
  = 50$ (red), $70$ (green) and $90$ (blue) from left to right.  The curves
  open to the right are contours for $\sin 2 \phi_s = 0.5$ (dotted), $-0.5$
  (solid) and the SM value $-0.07 \pm 0.01$ (dashed).
\label{fig:contour}}
\end{figure}

In Fig.~\ref{fig:contour}, we show the overlayed plot of the contours of fixed
$x_s$ and those of fixed $\sin 2 \phi_s$.  The shaded region in the center
shows the experimentally excluded points in the $\phi_L-\rho_L$ plane that
induce $x_s$ values smaller than $20.6$.  The hatched area corresponds to the
parameter space points that produce $x_s$ values falling within the $1 \sigma$
range of the SM value of $26.3$.  Contours for higher values of $x_s$ are also
shown.  The SM predicted $\sin2\phi_s \simeq -0.07 \pm 0.01$ would appear as
narrow bands around the $\sin 2\phi_s = -0.07$ curves.  Note that even if the
$x_s$ measurement turns out to be consistent with the SM expectation, it is
still possible that the new physics contributions, such as the $Z'$ model
considered here, can alter the $\sin 2 \phi_s$ value significantly.  It is
therefore important to have a clean determination of both quantities
simultaneously.  Once $x_s$ and $\sin 2\phi_s$ are extracted from $B_s$ decays,
one can determine $\rho_L$ up to a two-fold ambiguity and $\phi_L$ up to a
four-fold ambiguity, except for the special case with $\sin 2\phi_s \simeq 0$.

$\Delta\Gamma_s$ can be determined with a high sensitivity by measuring the
lifetime difference between decays into $CP$-specific states and into
flavor-specific states.  Using the $J/\psi \phi$ mode, simulations determine
\cite{Ball:2000ba} that the LHC can measure the ratio $\Delta\Gamma_s /
\Gamma_s$ with a relative error $\alt 10\%$ for an actual value around $-0.15$.
Tevatron simulations show that $\Delta\Gamma_s / \Gamma_s$ can be measured with
a statistical error of $\sim 0.02$.  For a sufficiently large $\rho_L$, the
$\cos\theta$ factor in Eq.~(\ref{eq:widthdiff}) increases from $-1$ at $\phi_L
= 0^\circ$ (mod $180^\circ$) to the maximum $1$ at $\phi_L = 90^\circ$ (mod
$180^\circ$).  We are left with the phase ranges $0^{\circ} \alt \phi_L \alt
30^{\circ}$, $60^{\circ} \alt \phi_L \alt 120^{\circ}$, and $150^{\circ} \alt
\phi_L \alt 180^{\circ}$ (mod $180^\circ$) where a $3\sigma$ determination of
$\Delta\Gamma_s$ can be made.

Once the right-handed $Z'$ couplings are introduced, we also have to include
the new $|\Delta B| = 2$ operators $O_1^{LR}$, $O_2^{LR}$, and $O^{RR}$ defined
in Eq.~(\ref{eq:ops}) into the effective Hamiltonian that contributes to the
$B_s$-$\lbar B_s$ mixing.  The matrix element of $O^{RR}$ is the same as that
of $O^{LL}$, while those of $O_1^{LR}$ and $O_2^{LR}$ are
\begin{eqnarray}
\label{eq:lrme1}
  &&
  \lan \lbar B_s | O_1^{LR} | B_s \ran
  = - \frac{4}{3} \left( \frac{m_{B_s}}{m_b(m_b) + m_s(m_b)} \right)^2
    m_{B_s}^2 f_{B_s}^2 B_1^{LR}(m_b) \\
  &&
  \lan \lbar B_s | O_2^{LR} | B_s \ran
  = 2 \left( \frac{m_{B_s}}{m_b(m_b) + m_s(m_b)} \right)^2
    m_{B_s}^2 f_{B_s}^2 B_2^{LR}(m_b)
\label{eq:lrme2}
\end{eqnarray}

For the $Z'$ coupling to right-handed currents, we define new parameters
$\rho_R$ and the associated weak phase $\phi_R$:
\begin{equation}
  \rho_R e^{i\phi_R} \equiv \frac{g_2 M_Z}{g_1 M_{Z'}} B_{sb}^R ~.
  \label{eq:pararight}
\end{equation}
At the $M_W$ scale, we have additional contributions to the effective
Hamiltonian due to the right-handed currents, similar to Eq.~(\ref{eq:para}).
The terms due to the left-right mixing are
\begin{equation}
  {\cal H_{\rm eff}^{Z'}} \supset
  \frac{G_F}{\sqrt{2}} \rho_L \rho_R e^{i(\phi_L+\phi_R)}
  (O_1^{LR}+O_1^{RL} \; , \; O_2^{LR}+O_2^{RL})
  \left( \begin{array}{c}
      1 \\ 0
    \end{array} \right ) ~.
\end{equation}
In the RGE running, the Wilson coefficient of $O_1^{LR}$ mixes with that of
$O_2^{LR}$; the relevant anomalous dimension matrices are \cite{Buras:2000if}
\begin{eqnarray}
  \gamma^{(0)} &=& \left( \begin{array}{cc}
    \frac{6}{N_c}&12 \\
    0 & -6 N_c + \frac{6}{N_c}
    \end{array}
        \right) ~, \\
  \gamma^{(1)} &=& \left( \begin{array}{cc}
    \frac{137}{6}+\frac{15}{2N_c^2}-\frac{22}{3N_c} f & \frac{200}{3} N_c
    -\frac{6}{N_c} -\frac{44}{3} f \\
    \frac{71}{4}+\frac{9}{N_c} - 2 f & -\frac{203}{6} N_c^2 +
    \frac{479}{6} + \frac{15}{2N_c^2} + \frac{10}{3} N_c f - \frac{22}{3
      N_c} f
    \end{array} \right) ~,
\end{eqnarray}
where $N_c$ is the number of colors and $f$ is the number of active quarks.  At
the scale of the $B$ meson masses, the value of $f$ is $5$.

We take $m_b(m_b) = 4.4$ GeV, $m_s(m_b) = 0.2$ GeV, and
$\Lambda^{(5)}_{\lbar{MS}} = 225$ MeV.  Following Eqs.~(\ref{u0vd}-\ref{jvs}),
we find the effective Hamiltonian terms for the operators $O_{1,2}^{LR}$ at
$m_b$ to be
\begin{equation}
  {\cal H_{\rm eff}^{Z'}} \supset
  \frac{G_F}{\sqrt{2}} \rho_R \rho_L e^{i(\phi_L+\phi_R)}
  (O_1^{LR}+O_1^{RL} \; , \; O_2^{LR}+O_2^{RL})
  \left( \begin{array}{c}
      0.930 \\
      -0.711
    \end{array} \right ) ~.
\label{eq:hammix}
\end{equation}
Note that there is no contribution of the operator $O_2^{LR}$ at the $M_W$
scale.  It is induced through the operator mixing in RGE running and actually
has an important effect at the $m_b$ scale, as one can see from its Wilson
coefficient in Eq.~(\ref{eq:hammix}).

\begin{figure}[t]
\centerline{\includegraphics[width=8cm]{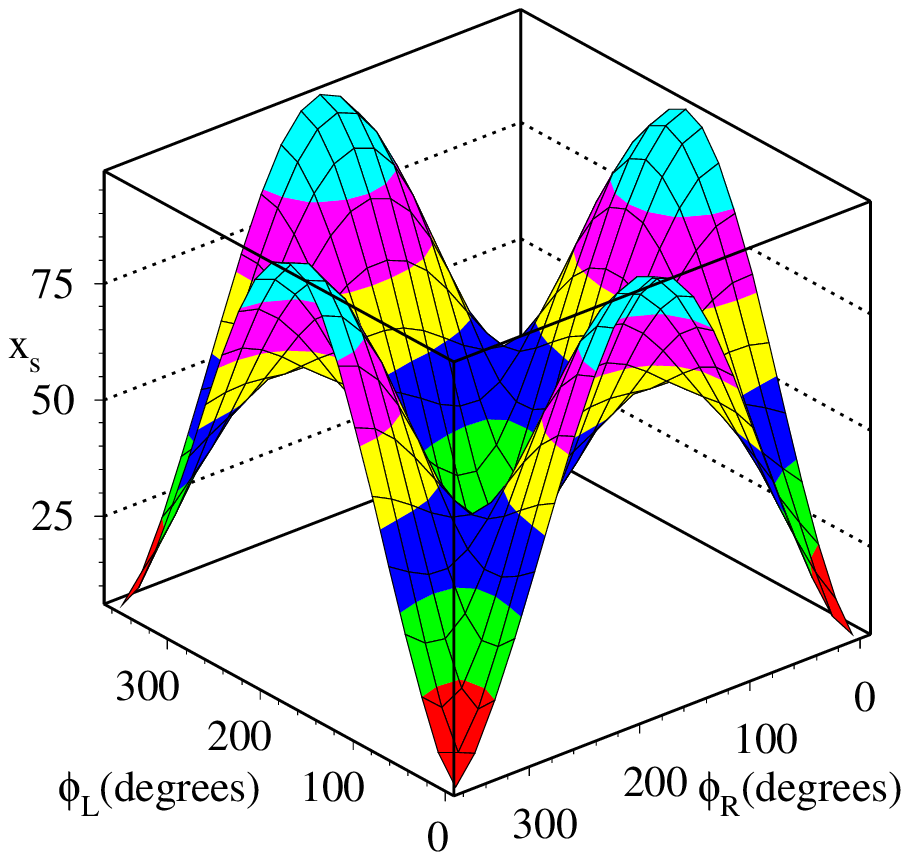}
            \includegraphics[width=8cm]{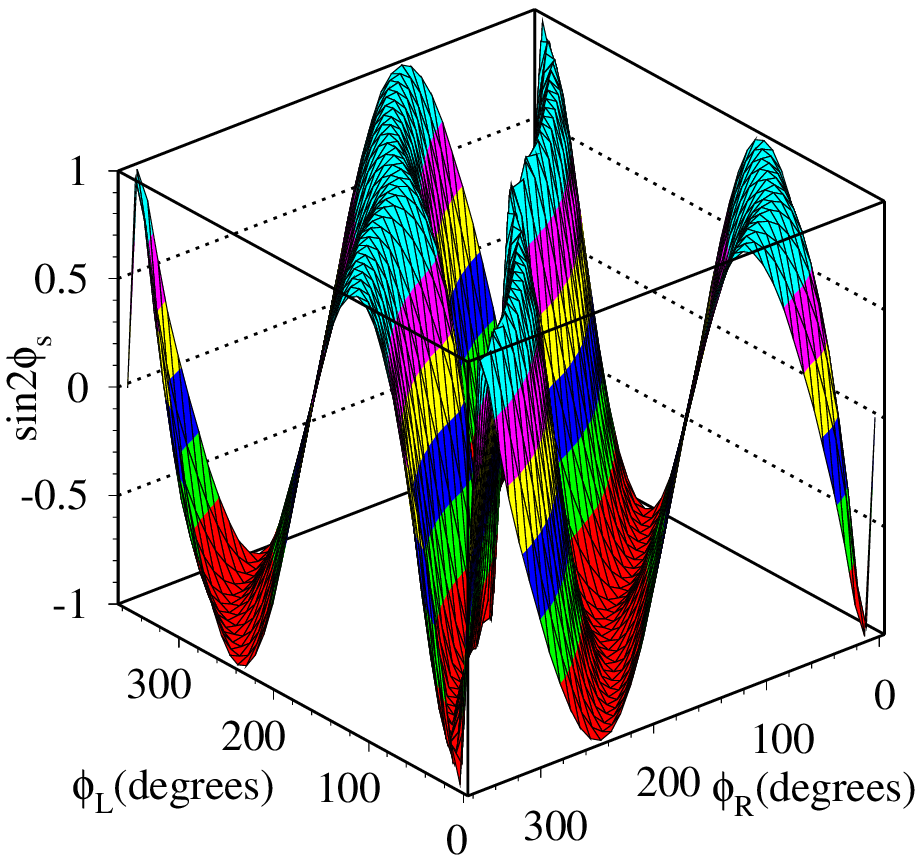}}
\centerline{(a)\hspace{8cm}(b)}
\caption[]{$x_s$ (a) and $\sin 2\phi_s$ (b) as functions of $\phi_L$
  and $\phi_R$ for $\rho_L = \rho_R = 0.001$.  The color shadings have no
  specific physical meaning.
\label{fig:all}}
\end{figure}

In the numerical analysis, we use the central values of $B_1^{LR}(m_b) = 1.753
\pm 0.021$ and $B_2^{LR}(m_b) = 1.162 \pm 0.007$ given in
Ref.~\cite{Becirevic:2001xt} with the decay constant $f_{B_s}$ the same as
before.  The predicted mass difference with all the $Z'$ contributions included
is then
\begin{eqnarray}
\Delta M_s = 18.0 \left| 1 + 3.858 \times 10^5 \left( \rho_L^2 e^{2
  i \phi_L} + \rho_R^2 e^{2 i \phi_R} \right) - % \nn \\
2.003 \times 10^6 \rho_L \rho_R e^{i (\phi_L + \phi_R)} \right| \;{\rm
  ps}^{-1}~. 
\end{eqnarray}
The overall contribution to $x_s$ from the SM and $Z'$ is
\begin{eqnarray}
  x_s = 26.3 \left| 1 + 3.858 \times 10^5 \left( \rho_L^2 e^{2 i \phi_L}
      + \rho_R^2 e^{2 i \phi_R} \right) - % \nn \\ 
  2.003 \times 10^6 \rho_L \rho_R e^{i (\phi_L + \phi_R)} \right| ~.
\label{eq:all}
\end{eqnarray}
To illustrate the interference among different contributions, we set $\rho_L =
\rho_R = 0.001$ and plot $x_s$ and $\sin 2\phi_s$ versus the weak phases
$\phi_L$ and $\phi_R$ in Fig.~\ref{fig:all} (a) and (b), respectively.

%Figure 3

First, we note that after the RGE running the operators $O_1^{LR}$ and
$O_2^{LR}$ interfere constructively.  This can be seen from the relative minus
sign between the Wilson coefficients in Eq.~(\ref{eq:hammix}) and a
corresponding relative minus sign in the hadronic matrix elements given in
Eqs.~(\ref{eq:lrme1}) and (\ref{eq:lrme2}).  Because of the constructive
interference and the fact that the bag parameters $B_1^{LR}$ and $ B_2^{LR}$
are twice as large as $B^{LL}$, the LR and RL operators together become the
dominant contributions.  The interference of all the terms makes $x_s$ reach
its maximum when one of the weak phases is $180^\circ$ and the other is
$0^\circ$ (mod $360^\circ$).  If $\rho_L$ and $\rho_R$ are both much smaller
than $10^{-3}$, the variation in $x_s$ in the $\phi_L$-$\phi_R$ space will be
indistinguishable from the predicted SM range.  Compared to Fig.~\ref{fig:ll3d}
(a) for $Z'$ with only $LL$ couplings, Fig.~\ref{fig:all} (a) shows that even
for large values of $\rho_L$ and $\rho_R$, $x_s$ can be smaller than $20.6$ due
to the interference among all the terms in Eq.~(\ref{eq:all}).  The current
$x_s \ge 20.6$ bound excludes the regions with $\phi_L + \phi_R \simeq 0^\circ$
(mod $360^\circ$). Because of the assumed equal values of $\rho_L$ and
$\rho_R$, the parameter space points with the same $\sin 2 \phi$ output lie
along directions that are roughly parallel to the $\phi_L + \phi_R = 360^\circ$
line.  For the more general cases of different $\rho_L$ and $\rho_R$ values,
the crests and troughs in Fig.~\ref{fig:all} (b) are no longer parallel to the
$\phi_L + \phi_R = 360^\circ$ line.

%%%%%%%%%%%%%%%%%%%%%%%%%%%%%%%%%%%%%%%%
\section{Conclusions \label{sec:conclusion}}
%%%%%%%%%%%%%%%%%%%%%%%%%%%%%%%%%%%%%%%%

In this paper we discuss the effects of a $Z'$ gauge boson with FCNC couplings
to quarks on the $B_s$-$\lbar B_s$ mixing.  We show how the mass difference and
$CP$ asymmetry are modified by the left-handed and right-handed $b$-$s$-$Z'$
couplings that may involve new weak phases $\phi_L$ and $\phi_R$.  In the
particular case of a left-chiral (right-chiral) $Z'$ model, one can combine the
measurements of $\Delta M_s$ (or $x_s$) and $\sin 2\phi_s$ to determine the
coupling strength $\rho_L$ ($\rho_R$) and the weak phase $\phi_L$ ($\phi_R$) up
to discrete ambiguities.  Once comparable left- and right-chiral couplings are
considered at the same time, we find the left-right interference terms dominate
over the purely left- or right-handed terms, partly because of the
renormalization running effects and partly because of the larger bag
parameters.

\vspace{0.2cm}

{\bf Acknowledgments:} C.-W.~C. would like to thank the hospitality of the high
energy theory group at University of Pennsylvania.  This work was supported in
part by the United States Department of Energy, High Energy Physics Division,
through Grant Contract Nos.\ DE-FG02-95ER40896, W-31-109-ENG-38, and
EY-76-02-3071.

%%%%%%%%%%%%%%%%%%%%%%%%%%%%%%%%%%%%%%%%

\end{document}